\documentclass{ws-p8-50x6-00}

\newcommand{\postscript}[2]{\setlength{\epsfxsize}{#2\hsize}
   \centerline{\epsfbox{#1}}}
\def\be{\begin{equation}}
\def\ee{\end{equation}}
\def\bear{\begin{eqnarray}}
\def\eear{\end{eqnarray}}

\def\NPB#1#2#3{{\it Nucl.~Phys.} {\bf{B#1}} (19#2) #3}
\def\PLB#1#2#3{{\it Phys.~Lett.} {\bf{B#1}} (19#2) #3}
\def\PRD#1#2#3{{\it Phys.~Rev.} {\bf{D#1}} (19#2) #3}
\def\PRL#1#2#3{{\it Phys.~Rev.~Lett.} {\bf{#1}} (19#2) #3}
\def\ZPC#1#2#3{{\it Z.~Phys.} {\bf C#1} (19#2) #3}
\def\PTP#1#2#3{{\it Prog.~Theor.~Phys.} {\bf#1}  (19#2) #3}

\def\EPJ#1#2#3{{\it Eur.~Phys.~J.} {\bf C#1} (19#2) #3}

\def\roughly#1{\raise.3ex\hbox{$#1$\kern-.75em\lower1ex\hbox{$\sim$}}}

\newcommand{\ms}{\mbox{\footnotesize$\overline{\rm MS}$~}}

\newcommand{\tAt}{X_t}

\newcommand{\tilt}{\tilde{t}}

\newcommand{\hxt}{\hat{X}_t}
\newcommand{\hyt}{\hat{Y}_t}

\newcommand{\mtilt}{m_{\tilde{t}}}

\newcommand{\gsim}{\lower.7ex\hbox{$\;\stackrel{\textstyle>}{\sim}\;$}}
\newcommand{\lsim}{\lower.7ex\hbox{$\;\stackrel{\textstyle<}{\sim}\;$}}

\begin{document}

\title{\vspace*{-2.4cm}\rightline
{\rm MADPH-00-1164}\rightline{\rm\lowercase 
{hep-ph/0003248}}\rightline{\rm M\lowercase{arch}, 2000}
\vspace{1cm}
Effective potential calculation of the MSSM lightest 
CP-even Higgs boson mass\thanks{\rm C\lowercase{ontribution to} PASCOS99:
\lowercase{7th} I\lowercase{nternational} 
S\lowercase{ymposium on} P\lowercase{articles}, S\lowercase{trings 
and} C\lowercase{osmology},
G\lowercase{ranlibakken}, T\lowercase{ahoe} 
C\lowercase{ity}, C\lowercase{alifornia}, 10-16 D\lowercase{ec} 1999.}}

\author{Ren-Jie Zhang}

\address{Department of Physics, University of Wisconsin, Madison WI 53706
USA}

\address{\small\it In memory of Prof. Xi-De Xie.}


\maketitle

\abstracts{
I summarize results of
two-loop effective potential calculations of the lightest CP-even
Higgs boson mass in the minimal supersymmetric standard model.}

Computing the lightest CP-even Higgs
boson mass is the most important loop calculation in the minimal
supersymmetric standard model because of the paramount importance
of a precise $m_{h^0}$ value to the Higgs boson experimental discovery.
Tree-level supersymmetry
relations require that the Higgs field quartic coupling be
related to the electroweak gauge couplings; 
therefore they impose a strict upper
bound $m_{h^0}\leq m_Z$, which is already 
in conflict with the current lower limit from LEP 2.

It is well-known that this tree-level limit can be drastically
changed by radiative corrections. One-loop calculations\cite{rad}
show that incomplete cancellations of the top and stop loops
give positive corrections of the form
\begin{equation}
\Delta m^2_{h^0} = {3h^2_t m^2_t\over 4\pi^2}
\ln{m^2_{\tilt}\over m^2_t}\ ,
\end{equation}
where $m_t$ and $m_{\tilt}$ are top and stop masses respectively.
This formula, however, suffers from an ambiguity in the definition of
$m_t$. Numerically, using running or on-shell top-quark
mass can amount to about 
$20\%$ difference in $\Delta m^2_{h^0}$. The problem can only be resolved 
by an explicit two-loop calculation.

Two-loop calculations in the existing literature have 
used two different approaches:
(a) a renormalization group resummation approach\cite{CEQWHHH}, and 
(b) a two-loop diagrammatic approach\cite{HH,HHW,RenJie}.
In the first approach, leading and next-to-leading logarithmic
corrections are calculated by integrating one- and two-loop renormalization
group equations.
However, two-loop non-logarithmic finite corrections are
not calculable in principle. The second approach 
was initiated by Hempfling and Hoang\cite{HH} using an effective
potential method;
they restricted their calculation to specific choice of supersymmetry 
parameters:
{\it i.e.} large $\tan\beta\rightarrow\infty$ and zero left-right stop mixing.
Two-loop QCD corrections were later computed at
more general cases\cite{RenJie} in the effective potential approach.
$m_{h^0}$ to the same two-loop QCD order was also computed\cite{HHW} 
using an explicit diagrammatic method. 
These calculations incorporate both two-loop logarithmic and 
non-logarithmic finite corrections. In the following, I shall
concentrate on the effective potential approach.

The general way of calculating corrections
to CP-even Higgs boson mass is to compute Higgs self-energy
and tadpole diagrams to the required loop order. 
In an effective potential approach,
these diagrams can be derived from a generating functional,
{\it i.e.} the effective potential, by taking explicit derivatives with 
respect to the Higgs fields. These quantities then enter the MSSM 
CP-even Higgs boson mass-squared matrix as follows
\begin{equation}
{\cal M}^2_h = \left[
\begin{array}{cc}
m^2_Z c_\beta^2 + m^2_{A^0} s^2_\beta + \Delta{\cal M}^2_{11} &
-(m^2_Z + m^2_{A^0}) s_\beta c_\beta + \Delta{\cal M}^2_{12} \\
-(m^2_Z + m^2_{A^0}) s_\beta c_\beta + \Delta{\cal M}^2_{21} &
m^2_Z s_\beta + m^2_{A^0} c_\beta + \Delta{\cal M}^2_{22}
\end{array}\right]\ ,\label{mhmatrix}
\end{equation}
where $\Delta{\cal M}^2_{ij}$ 
represents radiative corrections to the $ij$-entry.
We note that all these corrections are computed at the zero 
external momentum limit; sometimes it is necessary to calculate
self-energy diagrams directly to account for corrections from 
non-zero external momenta.

The CP-even Higgs boson masses can  be calculated by 
diagonalizing the above matrix in eq. (\ref{mhmatrix}). This computation
is tedious but can be greatly simplified 
when one considers the case $m_{A^0}\gg m_Z$,
where $m_{A^0}$ is the mass of the pseudoscalar $A^0$.
In this case, we find the corrections to $m^2_{h^0}$ is
\begin{equation}
\Delta m^2_{h^0} = {4 m^4_t\over v^2}\left({d\over d m^2_t}\right)^2 V
-{\rm Re}~\Pi_{hh}(m^2_{h^0})+{\rm Re}~\Pi_{hh}(0)\ .
\end{equation}
where $V$ is the effective potential, $v$ the Higgs field VEV,
and the last two terms account for non-zero external 
momentum corrections. 

We have carried out this calculation procedure to the two-loop order including
leading QCD\cite{RenJie} and top Yukawa\cite{EZ} corrections. To illustrate
our analysis, we present an approximation formula which
is derived under the following assumptions:
the soft masses for left and right stops, gluino, heavy Higgs bosons
and Higgsinos have a common
mass $M_S$, where $M_S$ can be identified as the supersymmetry scale.
The two eigenvalues and mixing angle of 
stops are then accordingly
$m^2_{{\tilt}_1}=\mtilt^2+m_t\tAt$,
$m^2_{{\tilt}_2}=\mtilt^2-m_t\tAt$ and
$s_t =c_t={1\over\sqrt{2}}$,
where the average top-squark mass $\mtilt^2=M_S^2+m_t^2$,
and $X_t=A_t+\mu/\tan\beta$ is the left-right stop mixing parameter.

We find the approximation formula for two-loop QCD+top Yukawa 
corrections is\cite{EZ} (in terms of on-shell mass parameters)
\begin{eqnarray}
&&\Delta m^2_{h^0} =
{3m^4_t\over2\pi^2 v^2}\biggl(\ln{\mtilt^2\over m_t^2}
+\hat{X}_t^2-{\hat{X}_t^4\over12}\biggr)\nonumber\\
&+&{\alpha_sm^4_t\over\pi^3 v^2}
\biggl(-3\ln^2{\mtilt^2\over m^2_t}-6\ln{\mtilt^2\over m^2_t}
+6\hat{X}_t-3\hat{X}_t^2\ln{\mtilt^2\over m^2_t}
-{3\hat{X}_t^4\over4}\biggr)\nonumber\\
&+&{3\alpha_t m_t^4\over 16\pi^3 v^2}\Biggl\{
s_\beta^2\biggl(3 \ln^2{M_S^2\over m_t^2} 
+ 13 \ln{M_S^2\over m_t^2}\biggr) -1 -{\pi^2\over3}
+c_\beta^2\biggl(
60 K + {13\over2} + {4\pi^2\over3}\biggr) \nonumber\\
&+&\biggl[3 s^2_\beta \ln{M_S^2\over m_t^2}
-c^2_\beta\biggl({69\over2}+24K\biggr)+41 
\biggr] \hxt^2
-\biggl(1+{61\over12}s^2_\beta\biggr)\hxt^4
+{s^2_\beta\over2}\hxt^6\nonumber\\
&+&c^2_{\beta}\biggl[(3-16 K-\sqrt{3}\pi)(4\hxt\hyt+\hyt^2)
+\biggl(16K+{2\pi\over\sqrt{3}}\biggr)\hxt^3\hyt\nonumber\\
&+&\biggl(-{4\over3}+24 K+\sqrt{3}\pi\biggr)\hxt^2\hyt^2
-\biggl({7\over12}+8 K+{\pi\over2\sqrt{3}}\biggr)\hxt^4\hyt^2\biggr]\Biggr\}\ ,
\label{eq:dmdr}
\end{eqnarray}
where the constant $K\simeq -0.195$. We note that two-loop QCD corrections
depend only on $\hat{X}_t=X_t/m_{\tilt}$ while the top Yukawa corrections
depend on $\hat{Y}_t=(A_t-\mu\tan\beta)/m_{\tilt}$ as well.
This approximation formula is good to a 
level of 0.5 GeV for most of the parameter space.

\begin{figure}[ht]
\postscript{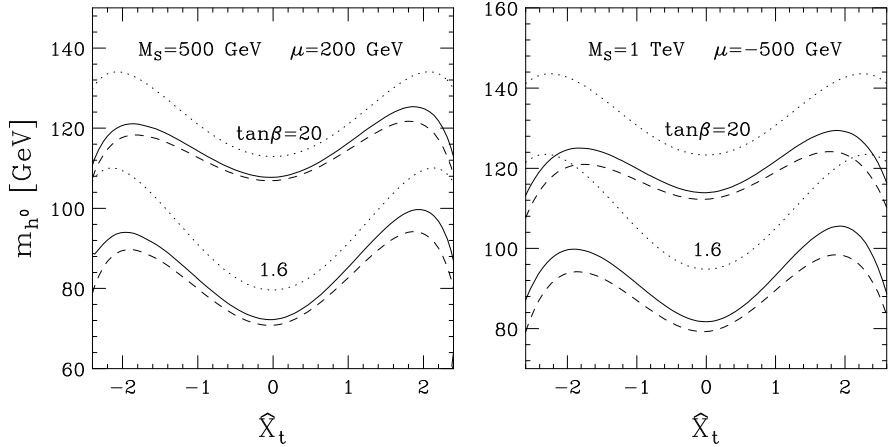}{1}
\caption{Higgs boson mass $m_{h^0}$ versus $\hat{X}_t$. 
The dotted, dot-dashed and solid lines correspond to Higgs boson masses 
calculated to the orders of one-loop, two-loop QCD and two-loop QCD+top Yukawa
respectively.}
\label{fig:1}
\end{figure}

Fig. 1 shows the Higgs boson mass $m_{h^0}$ vs. the stop
mixing parameter $\hat{X}_t$, for different choices of $M_S$, $\mu$ and
$\tan\beta$.
The two-loop
QCD corrections agree well with other approaches\cite{HHW}. 
They generally decrease $m_{h^0}$ from their one-loop values by $10-20$ GeV
depending on the parameter choice.
Two-loop Yukawa corrections are sizeable for large stop mixings, 
in particular, for $\hat{X}_t\simeq\pm 2$ two-loop
Yukawa corrections can increase $m_{h^0}$ by about $5$ GeV.

Another interesting feature observed in the literature\cite{HHW,RenJie}
is that
two-loop corrections shift the maximal mixing peaks. At the
one-loop level, these peaks are at $\hat{X}_t=\pm\sqrt{6}$. It is easy
to see from eq. (\ref{eq:dmdr}) that 
the size of shifts is about $10\%$,
{\it i.e.} the peaks move to $\hat{X}_t\simeq\pm 2$.
This is confirmed by Fig. 1.

Finally, renormalization group resummation technique can be 
used to derive a particularly nice mass correction formula
which has clearer physical interpretations. We find eq. (\ref{eq:dmdr})
can be transformed into the following form by using solutions to 
the renormalization group equations
\begin{equation}
\Delta m_{h^0}^2 =  
{3\overline{m}_t^4 (Q_t)\over 2\pi^2\overline{v}^2(Q^*_1)}
\ln{m^2_{\tilt}(Q_{\rm th})\over \overline{m}_t^2(Q_t')}
+{3\overline{m}_t^4 (Q_{\rm th})\over 2\pi^2\overline{v}^2(Q^*_2)}
\left[\hat{X}_t^2(Q_{\rm th})-{\hat{X}_t^4(Q_{\rm th})\over12}\right]
+\Delta^{(2)}_{\rm th}\ ,
\end{equation}
where $Q^*_1=e^{-1/3} m_t$, $Q^*_2=e^{1/3} m_t$,
$Q_t=\sqrt{m_t\mtilt}$, $Q_t'=(m_t\mtilt^2)^{1/3}$ and
$Q_{\rm th}=\mtilt$, $\overline{v}$ and $\overline{m}$ are the 
Standard Model $\ms$ parameters.
These choices of scales for evaluating one-loop corrections automatically
take into account two-loop leading and next-to-leading 
logarithmic effects. The leftover finite correction term 
$\Delta^{(2)}_{\rm th}$
is understood as two-loop threshold corrections and numerically small; its
detail form can be found in a forthcoming paper\cite{EZ}.

I thank J. R. Espinosa for collaborations.
This work was supported in part by a DOE grant 
No. DE-FG02-95ER40896 and in part by the
Wisconsin Alumni Research Foundation.

\end{document}